\documentclass[12pt]{article}

\usepackage{graphicx}
\begin{document}

\begin{center}
{\bf On a model of magnetically charged black hole with nonlinear electrodynamics } \\
\vspace{5mm} S. I. Kruglov
\footnote{E-mail: serguei.krouglov@utoronto.ca}
\underline{}
\vspace{3mm}

\textit{Department of Chemical and Physical Sciences, University of Toronto,\\
3359 Mississauga Road North, Mississauga, Ontario L5L 1C6, Canada} \\
\vspace{5mm}
\end{center}

\begin{abstract}
The Bronnikov model of nonlinear electrodynamics is investigated in general relativity. The magnetic black hole is considered and we obtain a solution giving corrections to
the Reissner-Nordstr\"{o}m solution. In this model spacetime at $r\rightarrow\infty$ becomes Minkowski's spacetime. We calculate the magnetic mass of the black hole and the metric function. At some parameters of the model there can be one, two or no horizons. The Hawking temperature and the heat capacity of black holes are calculated. We show that a second-order phase transition takes place and
black holes are thermodynamically stable at some range of parameters.
\end{abstract}

\section{Introduction}

The black hole (BH) physics is similar to ordinary thermodynamics. This analogy can help of understanding the theory of quantum gravity.
The semiclassical analysis of a BH radiation shows that the initial state information is hidden inside of the event horizon \cite{Hawking}.
Some questions were appeared on the information loss paradox and the unitary of the theory. To answer these questions a backreaction and quantum gravity effects should be taken into account. In addition, the semiclassical calculations do not work for light black holes. The effects
of quantum gravity can correctly describe the short distance behavior. At the same time, to avoid the short distance singularity, one can consider  regular black holes. A study of regular black holes without singularities allows us to treat
the minimum size as the Planck length. Not to search for quantum gravity, we concentrate here on the regular BH thermodynamics. Regular black hole solutions may be obtained by coupling general relativity and nonlinear electrodynamics (NLED).

Bardeen, Carter and Hawking \cite{bardeen} (see also \cite{Wald}) suggested that BH is a thermodynamic object which obeys four laws of black hole mechanics. Thus, we can study such thermodynamic properties of black holes as phase transitions, the thermal stability, the black hole evaporation and others.  Originally Davies and Hut \cite{Davies}, \cite{Hut} investigated BH phase transitions.
The BH thermodynamics phase transitions can be investigated with the aid of the heat capacity in the canonical ensemble. The BH is unstable if the heat capacity becomes negative. When the discontinuities of the heat capacity occur the phase transitions take place.

 The attractive feature of NLED is that an upper limit on the electric field at the origin of point-like particles takes place \cite{Born}-\cite{Kruglov1}.
In some models of NLED \cite{Born}-\cite{Kruglov2} the self-energy of charges is finite
when definite conditions for the Lagrangian are satisfied \cite{Shabad2}, \cite{Shabad3}.
Quantum corrections to classical electrodynamics also lead to NLED \cite{Heisenberg}-\cite{Adler}.
In addition, general relativity with NLED explains the universe inflation \cite{Garcia}-\cite{Kruglov3}. In the early universe the initial singularities also can be avoided in NLED models \cite{Novello1}.
In this paper we investigate BH solutions in the framework of Bronnikov's model of NLED \cite{Bronnikov}. The correspondence principle holds in this model because at the weak field limit NLED is transformed into Maxwell's electrodynamics. It was shown in \cite{Bronnikov} that the model includes the existence of a regular magnetic BH and monopoles. The theorem proven by \cite{Bronnikov} allows us to have a nontrivial case when the electric charge equals zero and the magnetic charge $q\neq 0$ so that the magnetic field is equal to $B=q/r^2$ and spacetime possesses a regular center. The entire mass of the BH has the electromagnetic origin.

The thermodynamics of black holes and phase transitions are studied in this paper. In \cite{Myung}, \cite{Breton_2} similar issues of the BH thermodynamics were discussed. Some aspects of black hole physics were studied in \cite{Bardeen}-\cite{Frolov}.

The paper is organized as follows. Field equations and energy-momentum tensor are described in section 2. In section 3 general relativity with NLED is studied. We obtain the asymptotic of the metric and mass functions at $r\rightarrow 0$ and $r\rightarrow\infty$. Corrections to the Reissner-Nordstr\"{o}m (RN) solution are found. In section 4 we calculate the Hawking temperature and the heat capacity of black holes. It was demonstrated that the second-order phase transition takes place in black holes. We find the range where black holes are stable. We made a conclusion in section 5.

The metric signature is given by $\eta=\mbox{diag}(-1,1,1,1)$ and we explore the units with $c=1$, $\varepsilon_0=\mu_0=1$.

\section{Field equations of NLED}

In this section we consider field equations in Minkowski's spacetime.
Let us study NLED, proposed in \cite{Bronnikov}, with the Lagrangian density
\begin{equation}
{\cal L} = -\frac{{\cal F}}{\cosh^2\sqrt[4]{|\beta{\cal F}|}},
 \label{1}
\end{equation}
where ${\cal F}=(1/4)F_{\mu\nu}F^{\mu\nu}=(\textbf{B}^2-\textbf{E}^2)/2$, the field tensor is defined as
$F_{\mu\nu}=\partial_\mu A_\nu-\partial_\nu A_\mu$ and the parameter $\beta$ has the dimension of (length)$^4$.
The modified model was proposed and investigated in \cite{Kruglov_3}.
At the weak field limit, $\beta {\cal F}\ll 1$, the Lagrangian density (1) becomes Maxwell's Lagrangian density, ${\cal L}\rightarrow-{\cal F}$, and as result, the correspondence principle holds.
Field equations found from Eq. (1), by the variation of the action corresponding to Lagrangian density (1) with respect to the 4-potential $A_\mu$, are given by
\begin{equation}
\partial_\mu\left({\cal L}_{\cal F}F^{\mu\nu} \right)=0,
\label{2}
\end{equation}
where
\begin{equation}\label{3}
 {\cal L}_{\cal F}=\partial {\cal L}/\partial{\cal F}=\frac{|\beta{\cal F}|^{1/4}\tanh\sqrt[4]{|\beta{\cal F}|}}{2\cosh^2\sqrt[4]{|\beta{\cal F}|}}-\frac{1}{\cosh^2\sqrt[4]{|\beta{\cal F}|}}.
\end{equation}
With the help of Eq. (1) one obtains the electric displacement field $\textbf{D}=\partial{\cal L}/\partial \textbf{E}$
\begin{equation}
\textbf{D}=\varepsilon\textbf{E},~~~~\varepsilon=-{\cal L}_{\cal F}.
\label{4}
\end{equation}
We find the magnetic field from the relation $\textbf{H}=-\partial{\cal L}/\partial \textbf{B}$,
\begin{equation}
\textbf{H}=\mu^{-1}\textbf{B},~~~~\mu^{-1}=-{\cal L}_{\cal F}=\varepsilon.
\label{5}
\end{equation}
Making use of Eqs. (4) and (5) one can represent field equations (2) as the Maxwell equations
\begin{equation}
\nabla\cdot \textbf{D}= 0,~~~~ \frac{\partial\textbf{D}}{\partial
t}-\nabla\times\textbf{H}=0.
\label{6}
\end{equation}
From the identity $\partial_\mu \tilde{F}^{\mu\nu}=0$, where $\tilde{F}^{\mu\nu}$ is the dual tensor, we obtain the second pair of nonlinear Maxwell's equations
\begin{equation}
\nabla\cdot \textbf{B}= 0,~~~~ \frac{\partial\textbf{B}}{\partial
t}+\nabla\times\textbf{E}=0.
\label{7}
\end{equation}
The relation, followed from Eqs. (4) and (5), is
\begin{equation}
\textbf{D}\cdot\textbf{H}=(\varepsilon^2)\textbf{E}\cdot\textbf{B}.
\label{8}
\end{equation}
Because $\textbf{D}\cdot\textbf{H}\neq\textbf{E}\cdot\textbf{B}$, according to the criterion of \cite{Gibbons}, the dual symmetry is broken.
In classical electrodynamics and in Born-Infeld electrodynamics the dual symmetry occurs but in QED, due to quantum corrections, the dual symmetry  is violated.

The symmetrical energy-momentum tensor can be found from the relation
\begin{equation}
T_{\mu\nu}=  {\cal L}_{\cal F}F_\mu^{~\alpha}F_{\nu\alpha}
-g_{\mu\nu}{\cal L}.
\label{9}
\end{equation}
From Eqs. (1), (3) and (9) one obtains the energy-momentum tensor trace
\begin{equation}\label{10}
 {\cal T}\equiv T^{\mu}_\mu=\frac{2{\cal F}\sqrt[4]{|\beta{\cal F}|}\sinh\sqrt[4]{|\beta{\cal F}|}}{\cosh^3\sqrt[4]{|\beta{\cal F}|}}.
\end{equation}
As ${\cal T}\neq 0$, the scale invariance is broken. At $\beta =0$ NLED (1) becomes Maxwell's electrodynamics, ${\cal T}=0$, and the scale invariance is recovered.

\section{Magnetized black holes}

The action of general relativity with NLED is given by
\begin{equation}
I=\int d^4x\sqrt{-g}\left(\frac{1}{2\kappa^2}R+ {\cal L}\right),
\label{11}
\end{equation}
where $\kappa^2=8\pi G\equiv M_{Pl}^{-2}$, $G$ is the Newton constant and $M_{Pl}$ is the reduced Planck mass.
The stability of a black hole with action (11) was studied in \cite{Breton_1}.
We investigate the magnetically charged black hole, and therefore, $\textbf{E}=0$, $\textbf{B}\neq 0$.
The Einstein and the electromagnetic field equations follow from action (11),
\begin{equation}
R_{\mu\nu}-\frac{1}{2}g_{\mu\nu}R=-\kappa^2T_{\mu\nu},
\label{12}
\end{equation}
\begin{equation}
\partial_\mu\left(\sqrt{-g}{\cal L}_{\cal F}F^{\mu\nu}\right)=0.
\label{13}
\end{equation}
We use the line element with the spherical symmetry
\begin{equation}
ds^2=-f(r)dt^2+\frac{1}{f(r)}dr^2+r^2(d\vartheta^2+\sin^2\vartheta d\phi^2),
\label{14}
\end{equation}
and the metric function is given by \cite{Bronnikov}
\begin{equation}
f(r)=1-\frac{2GM(r)}{r}.
\label{15}
\end{equation}
The mass function is defined as follows:
\begin{equation}
M(r)=\int_0^r\rho_M(r)r^2dr=\int_0^\infty\rho_M(r)r^2dr-\int_r^\infty\rho_M(r)r^2dr,
\label{16}
\end{equation}
with $\rho_M$ being the magnetic energy density and $ m_M = \int_0^\infty\rho_M(r)r^2dr$ is the magnetic mass of the black hole .
From Eq. (9) at \textbf{E}=0 we find the magnetic energy density
\begin{equation}\label{17}
  \rho_M=T_0^{~0}=-{\cal L} = \frac{{\cal F}}{\cosh^2\sqrt[4]{|\beta{\cal F}|}},
\end{equation}
where ${\cal F}=B^2/2=q^2/(2r^4)$, and $q$ is a magnetic charge.
It is convenient to introduce the dimensionless parameter $x=2^{1/4}r/(\beta^{1/4}\sqrt{q})$. Then using Eqs. (16) and (17) one obtains the mass function
\begin{equation}\label{18}
  M(x)=m_M-\frac{q^{3/2}}{2^{3/4}\beta^{1/4}}\tanh\left(\frac{1}{x}\right).
\end{equation}
The mass function presented by Eq. (18) has the same form as the solution given in
\cite{Bronnikov}.
We calculate the magnetic mass of the black hole
\begin{equation}\label{19}
 m_M = \int_0^\infty\rho_M(r)r^2dr=\frac{q^{3/2}}{2^{3/4}\beta^{1/4}}.
\end{equation}
Making use of Eqs. (15) and (18) one finds the metric function
\begin{equation}\label{20}
  f(x)=1-\frac{1-\tanh\left(1/x\right)}{bx},
\end{equation}
with $b=\sqrt{\beta}/(\sqrt{2}Gq)$.
With the help of Eq. (20) we obtain the asymptotic of the metric function at $r\rightarrow\infty$
\begin{equation}\label{21}
  f(r)=1-\frac{2Gm_M}{r}+\frac{Gq^2}{r^2}-\frac{G\sqrt{\beta}q^3}{3\sqrt{2} r^4}+\frac{G\beta q^4}{15 r^6}+
{\cal O}(r^{-8}).
\end{equation}
Equation (21) gives the corrections to the RN solution which are in the order of ${\cal O}(r^{-4})$.
At $r\rightarrow \infty$, $f(\infty)=1$, and the spacetime becomes flat.
It is easy to verify that
\begin{equation}\label{22}
  \lim_{x\rightarrow 0^+}f(x)=1.
\end{equation}
 Equation (22) shows that the black hole is regular without conical singularity.
The RN solution is recovered at $\beta=0$. The plot of the function $f(x)$  is represented in Fig. 1 for different parameters $b=\sqrt{\beta}/(\sqrt{2}Gq)$.
\begin{figure}[h]
\includegraphics[height=3.0in,width=3.0in]{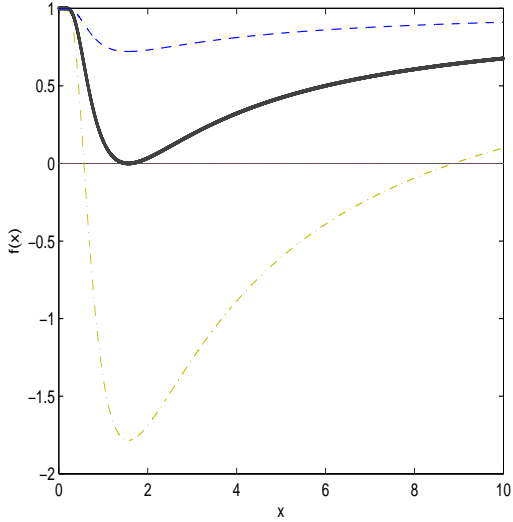}
\caption{\label{fig.1}The plot of the function $f(x)$. The dashed-dotted line corresponds to $b=0.1$, the solid line corresponds to $b=0.27846$ and the dashed line corresponds to $b=1$.}
\end{figure}
According to Fig. 1 at $b>0.27846$ there are no horizons. There is one horizon at $b\simeq 0.27846$ and the extreme singularity holds. When $b<0.27846$ we have two horizons. The horizons $x_h$ are defined by the equation $f(x_h)=0$. Then from Eq. (20) we come to the equation
\begin{equation}\label{23}
 b=\frac{1-\tanh(1/x_h)}{x_h}.
\end{equation}
The plot of the function $b(x_h)$ is represented in Fig. 2.
\begin{figure}[h]
\includegraphics[height=3.0in,width=3.0in]{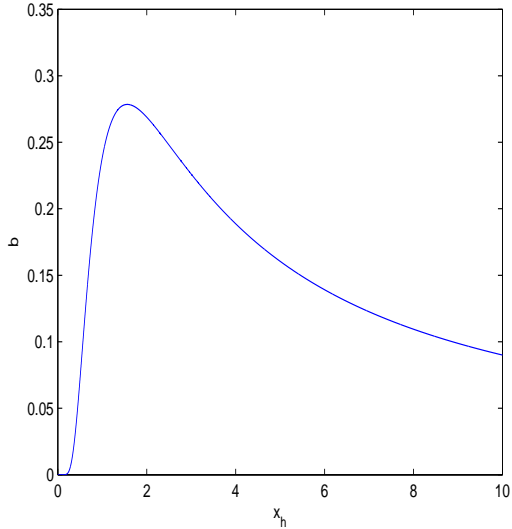}
\caption{\label{fig.2}The plot of the function $b(x_h)$.}
\end{figure}
From Eq. (23) we obtain the inner $x_-$ and the outer $x_+$ horizons of the black hole which are given in Table 1.
\begin{table}[ht]
\caption{The inner and outer horizons of the black hole}
\centering
\begin{tabular}{c c c c c c c c c c }\\[1ex]
 \hline
$b$ & 0.05 & 0.1 & 0.15 & 0.2 & 0.25 & 0.27 & 0.2784 \\[0.5ex]
\hline
$x_-$ & 0.4453 & 0.5654 & 0.6864 & 0.8347 & 1.0714 & 1.2680 & 1.5350 \\[0.5ex]
\hline
$x_+$ &18.9453 & 8.8784 & 5.4589 & 3.6705 & 2.4554 & 1.9702 & 1.5946 \\[0.5ex]
\hline
\end{tabular}
\end{table}
 It follows from Eqs. (10) at ${\cal F}=q^2/(2r^4)$ that at $r\rightarrow \infty$ the energy-momentum trace
becomes zero. As a result, in accordance with Eq. (12) the Ricci scalar $R=\kappa^2{\cal T}$ vanishes and spacetime becomes flat.

Let us discuss the equation of state of the BH inside the Cauchy horizon. From the expression for the pressure $p={\cal L}+(E^2-2B^2){\cal L}_{\cal F}/3$ and Eq. (17) we obtain the equation of state (for $E=0$) $p=-\rho_M-4{\cal F}{\cal L}_{\cal F}/3$. This equation holds for any radius $r$. At $r\rightarrow 0$, inside the Cauchy horizon, $p\rightarrow 0$, $\rho_M\rightarrow 0$, and ${\cal F}{\cal L}_{\cal F}\rightarrow 0$. As a result, the equation of state for a de Sitter spacetime $p=-\rho$ at $r\rightarrow 0$ is not the case for the BH described by the present model. But regular black holes proposed by \cite{Bardeen}, \cite{Hayward}, and \cite{Frolov} possess a de Sitter center. Thus, the geometry of the BH at $r\rightarrow0$ in Bronnikov's model is different compared to other models of regular black holes.

\section{Thermodynamics and phase transitions}

We will study the possible phase transitions and thermal stability of magnetized black holes. Let us calculate the Hawking temperature which is given by
\begin{equation}
T_H=\frac{\kappa_S}{2\pi}=\frac{f'(r_h)}{4\pi},
\label{24}
\end{equation}
where $\kappa_S$ is the surface gravity and $r_h$ is the horizon. Making use of Eqs. (15) and (16) we obtain the relations as follows:
\begin{equation}
f'(r)=\frac{2 GM(r)}{r^2}-\frac{2GM'(r)}{r},~~~M'(r)=r^2\rho_M,~~~M(r_h)=\frac{r_h}{2G}.
\label{25}
\end{equation}
It follows from Eq. (20) that
\[
f'(x)=\frac{1}{bx^2}-\frac{1}{bx^3\coth^2(1/x)}-\frac{\tanh(1/x)}{bx^2}.
\]
Then one can verify that indeed this equation leads to (25) taking into account $b=\sqrt{\beta}/(\sqrt{2}Gq)$, $x=2^{1/4}r/(\beta^{1/4}\sqrt{q})$, ${\cal F}=q^2/(2r^4)$ and Eqs. (17)-(19).
With the help of Eqs. (17) and (23) - (25), we find the Hawking temperature
\[
T_H=\frac{1}{2^{7/4}\pi\beta^{1/4}\sqrt{q}} \left(\frac{1}{x_h}-\frac{1}{x_h^2\left[1-\tanh(1/x_h)\right]\cosh^2(1/x_h)}\right)
\]
\begin{equation}
=\frac{1+b}{2^{7/4}\pi\beta^{1/4}\sqrt{q}x_h}\left(1-\frac{2}{(1+b)x_h}\right).
\label{26}
\end{equation}
The plot of the function $T_H(x_h)$ is given in Fig. 3.
\begin{figure}[h]
\includegraphics[height=3.0in,width=3.0in]{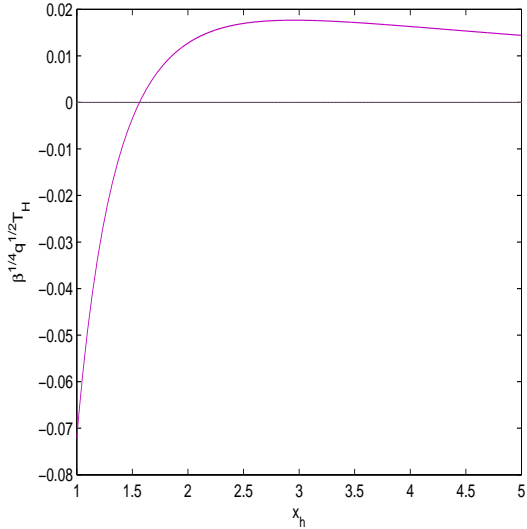}
\caption{\label{fig.3}The plot of the function $T_H\sqrt{q}\beta^{1/4}$ vs. horizons $x_h$.}
\end{figure}
The Hawking temperature becomes zero at $x_h\simeq 1.56$ ($b=0.27846$) corresponding to the extreme case. At $x_h>1.56$ the Hawking temperature is positive and the black hole is stable. When $x_h<1.56$ the Hawking temperature becomes negative and
the black hole is unstable. The maximum of the Hawking temperature holds at $x_+\simeq 2.95$ and
the heat capacity is singular. In this point the second-order phase transition takes place.
We calculate the heat capacity from the relation
\begin{equation}
C_q=T_H\left(\frac{\partial S}{\partial T_H}\right)_q=\frac{T_H\partial S/\partial r_h}{\partial T_H/\partial r_h}=\frac{2\pi r_hT_H}{G\partial T_H/\partial r_h}.
\label{27}
\end{equation}
The entropy obeys the Hawking area low $S=A/(4G)=\pi r_h^2/G$.
In Figs. 4 and 5 one can find the plots of the function $GC_q/(\sqrt{\beta}q)$ vs. the horizon $x_h$ for different values of $x_h$.
\begin{figure}[h]
\includegraphics[height=3.0in,width=3.0in]{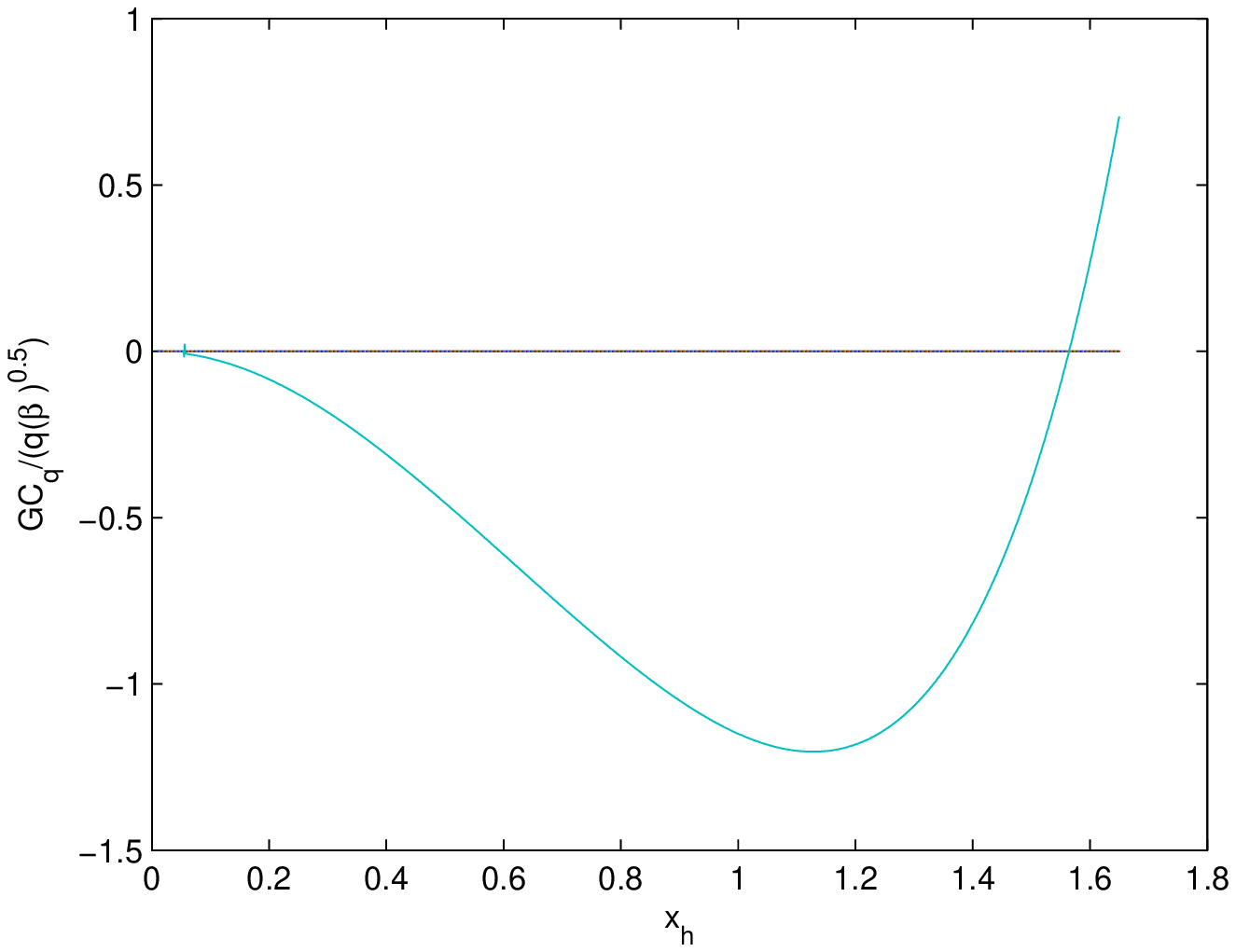}
\caption{\label{fig.4}The plot of the function $C_qG/(\sqrt{\beta}q)$ vs. $x_h$.}
\end{figure}
\begin{figure}[h]
\includegraphics[height=3.0in,width=3.0in]{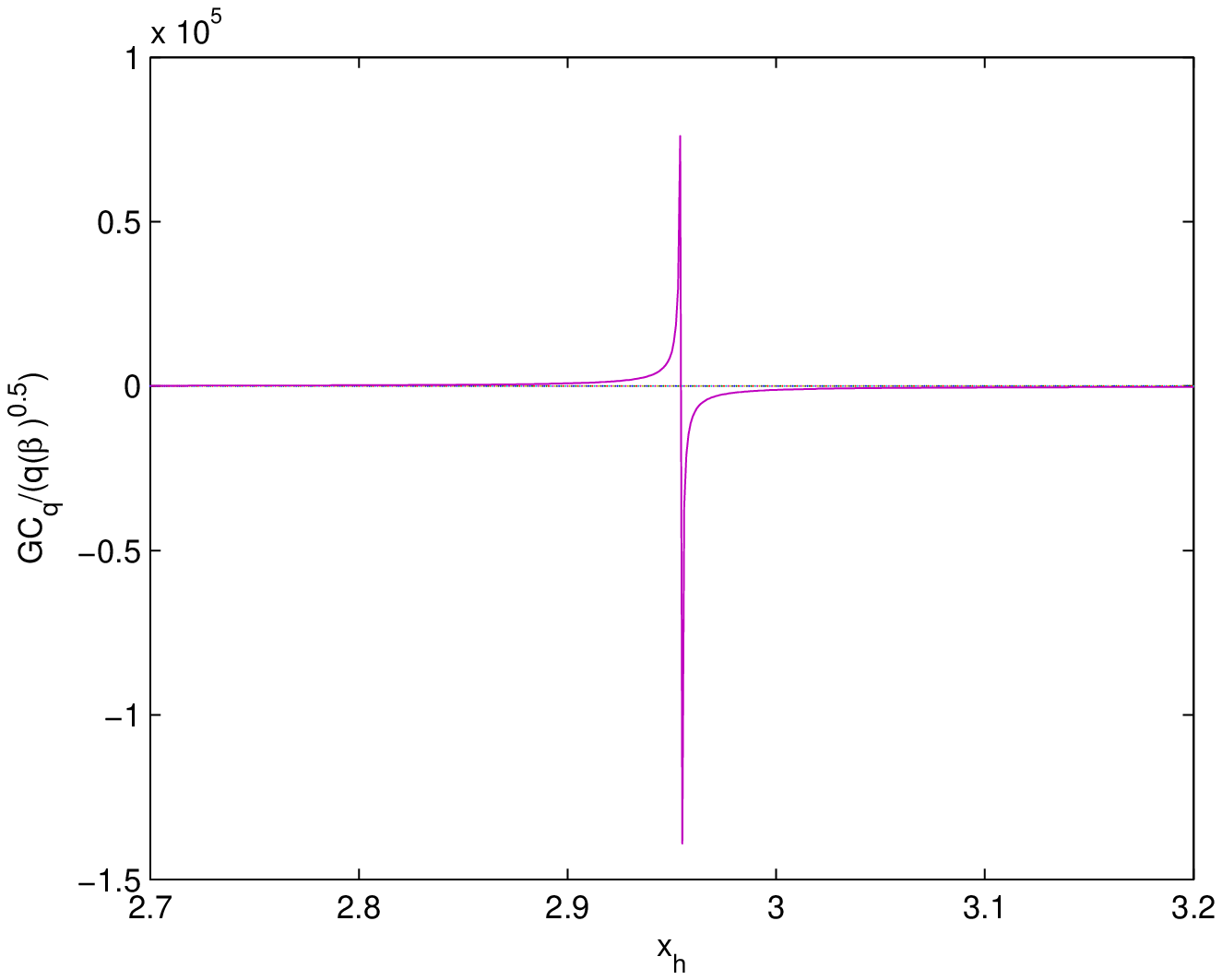}
\caption{\label{fig.5}The plot of the function $C_qG/(\sqrt{\beta}q)$ vs. $x_+$.}
\end{figure}
According to Figs. 4 and 5 the heat capacity possesses a discontinuity at $x_+\simeq 2.95$, and therefore,  the second-order phase transition of the black hole occurs. When $1.56\leq x_h<2.95$ the black hole is stable and at $x_+>2.95$ the heat capacity becomes negative and the black hole is unstable.

 \section{Conclusion}

We have investigated the Bronnikov model of NLED which at the weak field limit becomes Maxwell's electrodynamics and the correspondence principle holds. NLED coupled with the gravitational field was considered. The magnetized black holes were studied and we obtained the metric and the mass functions. At $r\rightarrow\infty$ we obtained corrections to the RN solution that are in the order of ${\cal O}(r^{-4})$.
Physical values of the theory depend on the parameter of NLED $\beta$.
The Hawking temperature and the heat capacity of black holes were calculated and we demonstrated that second-order phase transitions take place in black holes for definite parameters $\beta$.
The thermodynamic stabilities of black holes were studied and it was shown that in the range $1.56\leq x_+<2.95$  the black holes are stable.
In the framework of the non-commutative model with two horizons \cite{Arraut} similar phase transitions can happen. Phase transitions may appear also in alternative theories of gravity \cite{Arraut_0}.
It should be mentioned that in models of black holes with two horizons the Hawking temperature is connected with each horizons. This leads in some models to the existence of a minimal and maximal temperature in the black body radiation \cite{Nowakowski}.
Author of \cite{Arraut_1} found conditions in Massive Gravity for an observer to hold in the order to agree with the black-hole temperature.
Within the Generalized Uncertainty Principle authors of \cite{Nowakowski_1} shown that there can exist a black hole remnant with a mass $M_{Pl}$ corresponding to a maximal temperature $M_{Pl}$. There exists also a minimum length $l_{Pl}$.

\end{document}